\documentclass{article}
\usepackage{spconf,amsmath,graphicx}
\usepackage{hyperref}
\usepackage{tikz}
\usepackage{booktabs}
\usepackage[numbers,sort&compress]{natbib}
\usepackage{amssymb,xcolor}

\definecolor{myred}{RGB}{255,51,76}
\definecolor{myorange}{RGB}{251,139,35}
\definecolor{mygreen1}{RGB}{0,223,58}
\definecolor{mygreen2}{RGB}{0,115,30}

\definecolor{hnl}{RGB}{53, 117, 0}
\definecolor{hnr}{RGB}{204, 2, 32}
\definecolor{hnc}{RGB}{20, 20, 130}
\definecolor{ethr}{RGB}{196, 124, 252}
\definecolor{ethl}{RGB}{0, 190, 193}
\definecolor{ethc}{RGB}{72, 73, 77}

\title{Going Deeper With Brain Morphometry Using Neural Networks}
\name{
\begin{tabular}[t]{@{}c@{}}
  Rodrigo Santa Cruz$^{\star,\dag}$, L\'eo Lebrat$^{\star,\dag}$, Pierrick Bourgeat$^{\star}$, Vincent Dor\'e$^{\ast}$, Jason Dowling$^{\star}$, \\
  Jurgen Fripp$^{\star}$, Clinton Fookes$^{\dag}$, Olivier Salvado$^{\dag,\ddag}$ \thanks{Thanks to Maxwell plus~\url{http://maxwellplus.com/}.}
\end{tabular}}

\address{$^{\star}$ CSIRO Health and Biosecurity, The Australian eHealth Research Centre,\\
        $^{\dag}$Image and Video Laboratory QUT, $^{\ast}$ Department of Nuclear Medicine and Centre for PET. \\
        $^{\ddag}$ CSIRO Data61, Australia.}

\begin{document}
%
\maketitle
\begin{abstract}
Brain morphometry from magnetic resonance imaging (MRI) is a consolidated biomarker for many neurodegenerative diseases. 
Recent advances in this domain indicate that deep convolutional neural networks can infer morphometric measurements within a few seconds.
Nevertheless, the accuracy of the devised model for insightful bio-markers (mean curvature and thickness) remains unsatisfactory.
In this paper, we propose a more accurate and efficient neural network model for brain morphometry named \textit{HerstonNet}.
More specifically, we develop a 3D ResNet-based neural network to learn rich features directly from MRI, design a multi-scale regression scheme by predicting morphometric measures at feature maps of different resolutions, and leverage a robust optimization method to avoid poor quality minima and reduce the prediction variance.
As a result, \textit{HerstonNet} improves the existing approach by 24.30\% in terms of intraclass correlation coefficient (agreement measure) to FreeSurfer silver-standards while maintaining a competitive run-time.
\end{abstract}

\section{Introduction}
\label{sec:intro}


Brain morphometry from MRI is a non-invasive measurement of brain structures which is important for clinical studies of many neurodegenerative diseases~\citep{pettigrew2016cortical,nopoulos2007morphology}.
There exist well-established software for brain morphometry~\cite{Dahnke:NI2013:CAT,Kim:NI2005:CLASP,Shattuck:MIA2002:BrainSuite,Kriegeskorte:NI2001:BrainVoyage,Fischl:NI1999:freesurfer,henschel2020:fastsurfer}.
However, these tools rely on the construction of complex representations such as segmentation masks, partial volume maps, and triangular meshes of genus zero that are computationally expensive to obtain. Consequently, classical methods cannot be employed in healthcare applications where fast results are critical. 
For instance, \emph{FreeSurfer}~\citep{Fischl:NI1999:freesurfer} takes around \emph{10 hours} per scan, where its deep learning powered version \emph{FastSurfer}~\citep{henschel2020:fastsurfer} takes \emph{1.7 hours}.

Recently, \citet{rebsamen2020brain} reformulate brain morphometry as a regression problem. 
While this approach cuts down the brain morphometry processing time for computing the volume, mean thickness, and mean curvature of $165$ anatomical zones to a few seconds. It exhibits feeble results for thickness and curvature: only 38\% of the thickness measurements reach a good intraclass correlation coefficient (ICC $\geq$ 0.6), and 48\% of the curvature measurements present a very poor ICC to FreeSurfer measurements (ICC $<$ 0.4).

\begin{table*}[t]
\resizebox{\textwidth}{!}{\begin{tabular}{l cccc c cccc c cccc c cccc }
 & \multicolumn{4}{c}{Healthy Control} &  & \multicolumn{4}{c}{Mild Cognitive Impairment} &  & \multicolumn{4}{c}{Alzheimer's Disease} &  & \multicolumn{4}{c}{Other} \\ \cmidrule(l){2-5}\cmidrule(l){7-10}\cmidrule(l){12-15}\cmidrule(l){17-20}
  & S & N & Mean Age ($\pm$ SD)  & \% Male  &  & S & N & Mean Age ($\pm$ SD)  & \% Male  &  & S & N & Mean Age ($\pm$ SD)  & \% Male  &  & S & N & Mean Age ($\pm$ SD)  & \% Male   \\
\hline
Train & 441 & 1127 & 74.79 $\pm$ 6.42 & 44.99 & & 996 & 1972 & 73.98 $\pm$ 6.90 & 41.13 & & 537 & 922 & 76.30 $\pm$ 5.72 & 30.80 & & 1013 & 1611 & 72.74 $\pm$ 7.25 & 50.35 \\
Validation & 113 & 284 & 74.02 $\pm$ 7.35 & 27.82 & & 243 & 487 & 74.03 $\pm$ 6.16 & 43.12 & & 157 & 268 & 76.74 $\pm$ 6.23 & 28.73 & & 242 & 352 & 73.27 $\pm$ 7.55 & 55.32 \\
Test & 190 & 482 & 73.58 $\pm$ 6.37 & 50.62 & & 391 & 736 & 73.85 $\pm$ 6.23 & 40.76 & & 252 & 420 & 76.16 $\pm$ 6.27 & 22.38 & & 421 & 649 & 72.65 $\pm$ 7.22 & 56.67 \\
\hline
Overall & 744 & 1893 & 74.37 $\pm$ 6.57 & 43.85 & & 1630 & 3195 & 73.96 $\pm$ 6.64 & 41.35 & & 946 & 1610 & 76.34 $\pm$ 5.95 & 28.26 & & 1676 & 2612 & 72.80 $\pm$ 7.28 & 52.69 \\
\hline
\end{tabular}}
\caption{Demographic information of the ADNI+AIBL dataset. S and N denotes the number of subjects and data points respectively. ``Other'' column aggregates the subjects whose diagnosis is under-represented or not informed.}
\label{tab:data_stats}
\end{table*}

This paper proposes a novel neural network regression model for brain morphometry from MRI named \emph{HerstonNet}. 
The devised model leverages a ResNet-derived architecture~\citep{he2016deep}, which allows us to increase the depth of our model with limited overhead.
Since the morphometric problem focuses on structure with different resolutions, we employ a multiple regression ``heads'' approach to capture coarse to fine-grained details.
Finally, as training such a model is a challenging task, we introduce a modern optimization strategy that allows us to decrease the prediction variance while avoiding bad local minimum and overfitting.
As a result, \textit{HerstonNet} improves the intraclass correlation by 6.09\% for volumes, 21.73\% for thickness, and 43.15\% for the mean curvature.

\section{Materials and Methods}
\label{sec:methods}

\subsection{Data}
In this work, our data aggregates two public datasets: the Alzheimer’s Disease Neuroimaging Initiative (ADNI)~\cite{Jack2008:ADNI} and the Australian Imaging, Biomarkers and Lifestyle (AIBL) study~\cite{rowe2010amyloid}. 
These consist of T1-weighted MRIs and their silver-standard brain morphometric measurements are obtained with the FreeSurfer V6 pipeline~\cite{Fischl:NI1999:freesurfer}.
Every MRI is associated with 29 subcortical structure volumes, and 68 mean thickness and mean curvatures for the parcellations defined by the Desikan-Killiany (DK) atlas~\cite{desikan2006automated}.
We split this dataset into three folds, training ($\approx 60\%$), validation ($\approx 15\%$), and test ($\approx 25\%$); whose demographic statistics are described in Table~\ref{tab:data_stats}. 
These splits are patient-wise disjoint for an unbiased evaluation.  Note that this dataset is \emph{16 times} larger than the one described by~\citet{rebsamen2020brain}, which is key to train larger models and to obtain robust statistical estimates of their performance.


As MRI prepossessing, we perform re-sampling to 1$mm^3$ voxel size, non-uniform intensity normalization, and skull-striping using FreeSurfer's $\texttt{recon-all -autorecon1}$ tool. We also crop the final MRI volumes at the dimension of the largest skull-stripped brain resulting in MRI inputs of dimensions $164 \times 172 \times 202$.
Furthermore, during training, we rescale the target morphometric measurements between 0 and 1 using min-max normalization with the min and max taken along the entire training data. This normalization helps us to stabilize the early stages of the optimization.

\subsection{\textit{HerstonNet}}
Increasing the capacity and the depth of a neural network is key to achieve a strong performance~\cite{hornik1991approximation,huang2019gpipe}. In the former paper~\cite{rebsamen2020brain}, the authors proposed a VGG-type architecture. Trying to increase the size of this particular architecture, one faces two problems.
First and foremost, increasing the size of VGG results in longer training times and a more complicated optimization process~\cite{glorot2010understanding};
Secondly, once you pass a certain depth, a VGG-type architecture yields less accurate results (see~\cite{he2015convolutional} and discussion therein). This intuition has been confirmed recently by analyzing the landscape of these neural networks~\cite{li2018visualizing}. 
For those aforementioned reasons, we decide to adopt a ResNet-type~\cite{he2016deep} architecture allowing us to increase the depth of our network while keeping the training achievable. The main feature of this architecture is the so-called identity trick that allows circumventing the problem of vanishing gradients. 

\begin{figure*}[t!]
    \centering
    \includegraphics[width=500px,height=150px]{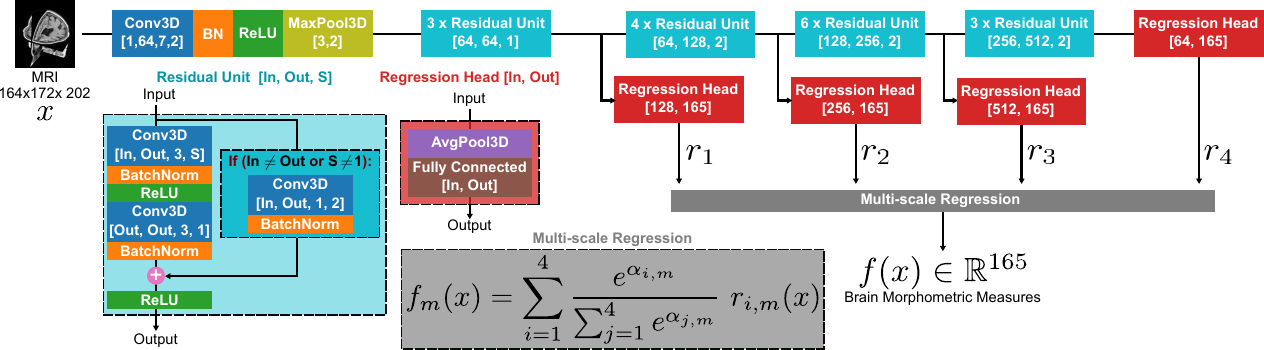}
    \caption{\textit{HerstonNet} Architecture (3D ResNet) with 4 regression ``heads''. 
    The layers are parametrized as follows: $\texttt{Conv3D[In, Out, k, s]}$ is a 3D convolutional layer with $\texttt{In}$ channel inputs, $\texttt{Out}$ channel outputs, $\texttt{k}^3$ kernel size, and $\texttt{S}$ convolution stride, $\texttt{MaxPool3D[k, s]}$ is a 3D Max Pooling layer with $\texttt{k}^3$ kernel size and $\texttt{S}$ convolution stride, and $\texttt{FullyConnected[In, Out]}$ is a fully connected layer with $\texttt{In}$ input channels and $\texttt{Out}$ output channels. Multi-scale regression is accomplished by the weighted average of the outputs of the four regression ``heads'' where the weights $\alpha \in \mathbb{R}^{4\times165}$ are also learnable parameters.}
    \label{fig:herstonnet_arch}
\end{figure*}

By its very nature, brain morphometry depends on information at different resolution. While volume measurements can be inferred from a low resolution, thickness and curvature require high-frequency information. 
Classical convolutional neural networks tend to average information along with its depth so that it is very involved for a funnel-like architecture to achieve good results for both volume and thickness/curvature. 
To circumvent this problem, we design a multiple regression ``heads'' scheme where these measures are predicted at different feature maps for different resolutions providing better accuracy to our model. The results of different resolutions are averaged with weights learned by the network, (note that the canonical simplex constraint is simply enforced using the softmax function). The resulting model is presented with further detail in Figure~\ref{fig:herstonnet_arch}.

\subsection{Optimization}

We train our neural network to minimize the mean squared error on batches of six MRIs using Adam optimizer with an initial learning rate of $10^{-4}$. 
As data-augmentation, we constrain ourselves to Gaussian noise injection, translations ($\pm15~voxels$), and small rotations ($\pm 30^{\circ}$) of the input MRIs. Note that more complex geometric transformations could invalidate the silver-standard morphometric measures (labels) used to train our model.
We also use early stopping by training our model for 170 epochs, periodically evaluating its performance on the validation set, and keeping the weights with the highest mean ICC across all of the measures in the validation set.

However, we notice that this approach leads to limited generalization properties, which translate into degraded confidence intervals of the ICC.
Inspired by the work of~\cite{izmailov2018averaging}, where the authors show that conventional stochastic optimization methods are struggling to reach the central point of a flat minimum, we take advantage of the SWA optimization method~\citep{izmailov2018averaging}.
The intuition behind this algorithm is to make use of a cyclical learning rate schedule after the initial training to find several stationary points around a wide minimum. Then we average the weights of all the stationary points found so far to land in the central part of the wide minimum. In our experiments, this averaged point benefits from a better generalization property.
Therefore, after training the model with Adam, we apply SGD for five cycles using a cyclic linear learning rate schedule from $10^{-2}$ to $10^{-6}$ during four epochs. The final model consists of the average of the weights at the minimal learning rate values. 
SWA optimization improves the mean ICC by $1.99\%$,  and reduces its confidence intervals by $35.59\%$.

\subsection{Evaluation}

\begin{figure}[hb!]
    \centering
    \includegraphics[width=0.48\textwidth]{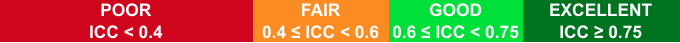}
    \vspace{-20px}
    \caption{ICC intervals for clinical applications according to \citet{cicchetti1994guidelines}.}\label{fig:ViadoUsingNegativeSpaces}
    \vspace{-10px}
    \vspace{-20px}
\end{figure}

In our experiments, we compare the proposed model with the previous work by  \citet{rebsamen2020brain} on brain morphometry from MRI.
We employ the intraclass correlation coefficient (ICC) along with a 95\% confidence interval as the evaluation metric. 
Following the research reliability guidelines for reporting ICC values~\citep{koo2016guideline}, we compute the two-way mixed effects, absolute agreement, single rater (ICC(2,1)) between the predicted brain morphometric measurements and the FreeSurfer silver-standard used as ground-truth. 
This metric produces negative values for negative correlation, zero for no correlation like just predicting the average of a measurement, and one for a perfect correlation between the predicted and ground-truth values.
Furthermore, as discussed by \citet{cicchetti1994guidelines} and used in \citep{rebsamen2020brain}, we make use of the ICC intervals routinely used for clinical applications and described in Figure~\ref{fig:ViadoUsingNegativeSpaces}.

\begin{figure}[h!]
    \includegraphics[width=.49\textwidth]{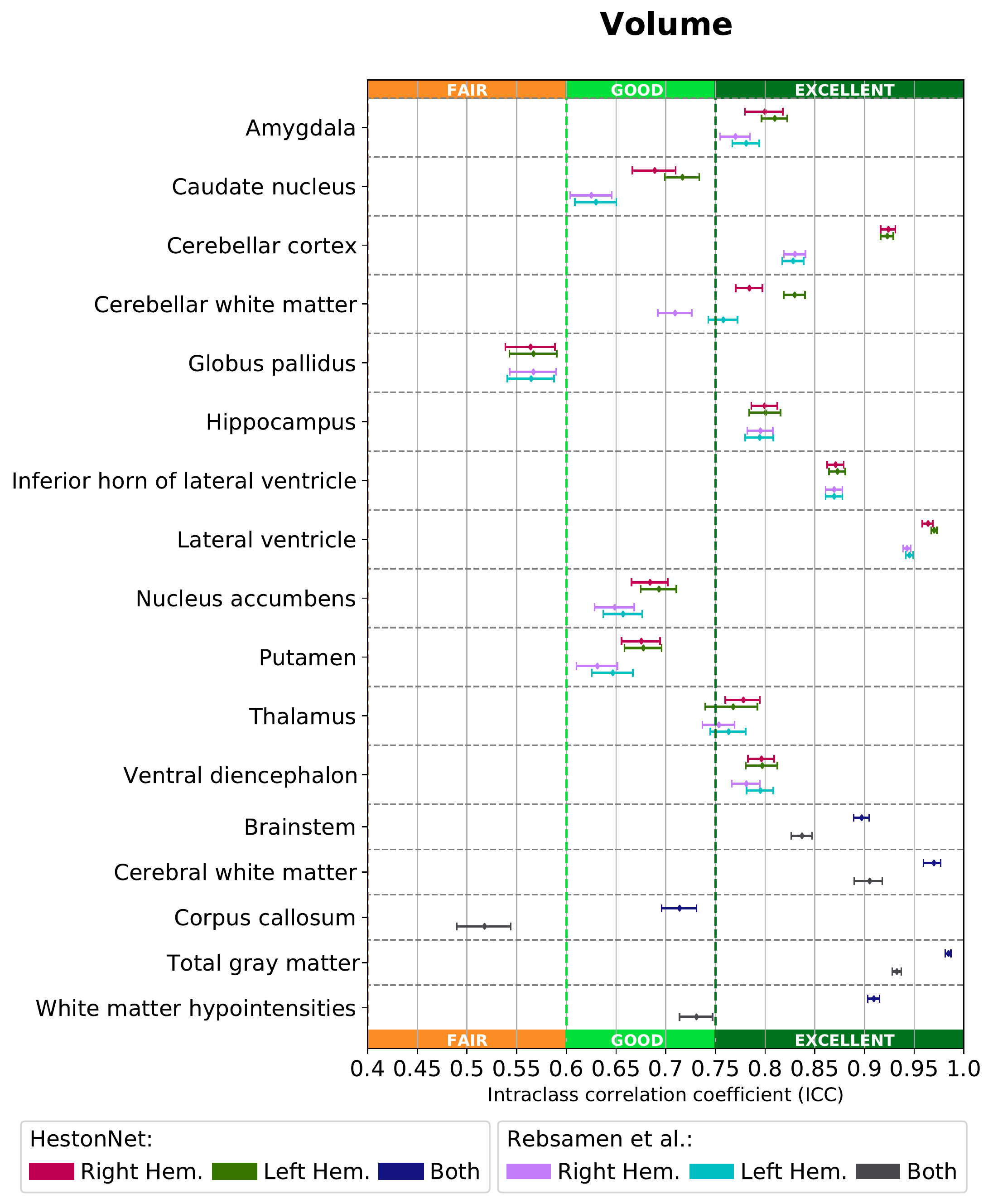}
    \vspace{-20px}
    \caption{ICC with 95\% confidence intervals for the volumes of 29 subcortical structures.}
    \label{fig:volumes}
    \vspace{-20px}
\end{figure}

\section{Results}

\begin{figure*}[!htb]
    \centering
    \includegraphics[height=0.82\textheight]{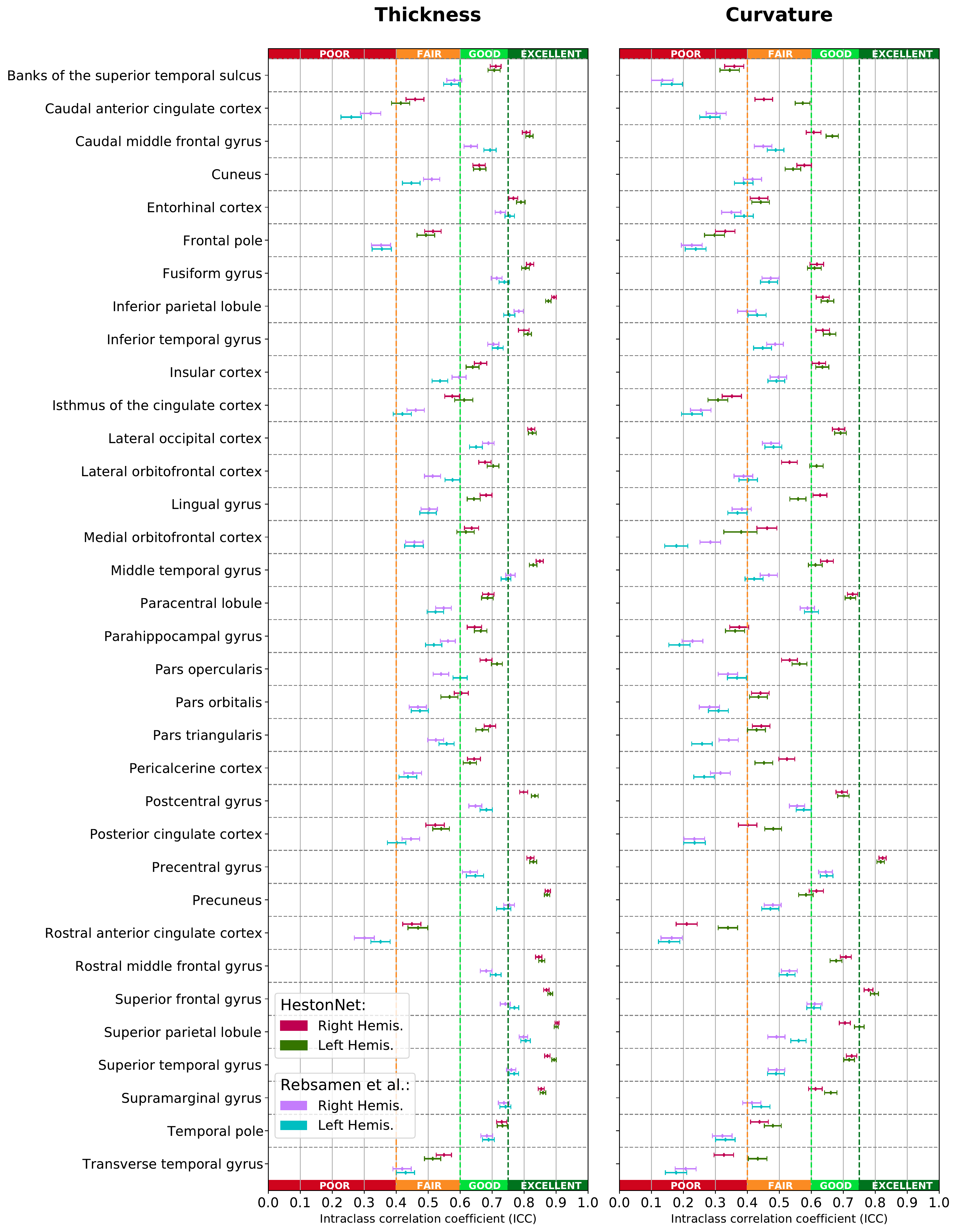}
    \vspace{-10px}
    \caption{ICC with 95\% confidence intervals for the thickness and curvature of 68 cortical parcellations.}
    \label{fig:thk_curv}
\end{figure*}

In Figures \ref{fig:volumes} and \ref{fig:thk_curv} we present the ICC values along with 95\% confidence intervals for HestonNet and the baseline model~\citep{rebsamen2020brain} for the predictions of 29 volumes, and 68 cortical thickness and curvature.
In summary, HestonNet achieves the mean ICC of $0.665 (\pm 0.166)$, while the baseline presents mean ICC of $0.535 (0.192)$ providing an overall improvement of $24.30\%$. 
It is also important to highlight that our model has comparable time complexity. It can process a batch of 6 MRIs in just 2.10 seconds while the baseline spends 1.93 seconds. 
Thus, \textit{HerstonNet} is a more accurate and still very efficient method for brain morphometry from MRI.

We observe that both methods present satisfactory results in terms of mean ICC for predicting volumes of subcortical structures ($0.755~vs~0.801$), while their performances differ significantly for thickness ($0.589~vs~0.717 $) and curvature estimation ($0.387~vs~0.554$).
Considering only thickness measures, our method achieves a good ICC on 56 out of 68 parcellations, while the baseline method achieves this level for only 33 parcellations.
Furthermore, the baseline presents very poor performance for the thickness of 6 parcellations, while our method does not present any poor ICC. 
When considering curvatures, the difference is even greater, our method's mean ICC is $43.15\%$ better than the baseline. Moreover, \textit{HerstonNet} presents poor ICC for only 12 out of 68 parcellations, while the baseline method~\cite{rebsamen2020brain} presents poor parcellation for 41 out of 68 parcellations.

\section{Conclusion}
\label{sec:con}

In this paper, we present \textit{HerstonNet} a 3D ResNet architecture with multi-scale regressors, trained with an advanced optimization technique. In our experiments, we show that \textit{HerstonNet} is a more accurate and still efficient solution for brain morphometry from MRI. \textit{HerstonNet} provides impressive thickness and curvature estimation precision, which has a pivotal role in neurodegenerative disease diagnosis.

\section{Compliance with Ethical Standards}
This research was approved by CSIRO ethics 2020\_068\_LR.

\bibliographystyle{IEEEtranN}
\bibliography{long,biblio}

\begin{thebibliography}{21}
\providecommand{\natexlab}[1]{#1}
\providecommand{\url}[1]{#1}
\csname url@samestyle\endcsname
\providecommand{\newblock}{\relax}
\providecommand{\bibinfo}[2]{#2}
\providecommand{\BIBentrySTDinterwordspacing}{\spaceskip=0pt\relax}
\providecommand{\BIBentryALTinterwordstretchfactor}{4}
\providecommand{\BIBentryALTinterwordspacing}{\spaceskip=\fontdimen2\font plus
\BIBentryALTinterwordstretchfactor\fontdimen3\font minus
  \fontdimen4\font\relax}
\providecommand{\BIBforeignlanguage}[2]{{%
\expandafter\ifx\csname l@#1\endcsname\relax
\typeout{** WARNING: IEEEtranN.bst: No hyphenation pattern has been}%
\typeout{** loaded for the language `#1'. Using the pattern for}%
\typeout{** the default language instead.}%
\else
\language=\csname l@#1\endcsname
\fi
#2}}
\providecommand{\BIBdecl}{\relax}
\BIBdecl

\bibitem[Pettigrew et~al.(2016)Pettigrew, Soldan, Zhu, Wang, Moghekar, Brown,
  Miller, Albert, Team, et~al.]{pettigrew2016cortical}
C.~Pettigrew, A.~Soldan, Y.~Zhu, M.-C. Wang, A.~Moghekar, T.~Brown, M.~Miller,
  M.~Albert, B.~R. Team \emph{et~al.}, ``Cortical thickness in relation to
  clinical symptom onset in preclinical ad,'' \emph{NeuroImage: Clinical},
  vol.~12, pp. 116--122, 2016.

\bibitem[Nopoulos et~al.(2007)Nopoulos, Magnotta, Mikos, Paulson, Andreasen,
  and Paulsen]{nopoulos2007morphology}
P.~Nopoulos, V.~A. Magnotta, A.~Mikos, H.~Paulson, N.~C. Andreasen, and J.~S.
  Paulsen, ``Morphology of the cerebral cortex in preclinical huntington’s
  disease,'' \emph{American Journal of Psychiatry}, vol. 164, no.~9, pp.
  1428--1434, 2007.

\bibitem[Dahnke et~al.(2013)Dahnke, Yotter, and Gaser]{Dahnke:NI2013:CAT}
R.~Dahnke, R.~A. Yotter, and C.~Gaser, ``Cortical thickness and central surface
  estimation,'' \emph{NeuroImage}, vol.~65, pp. 336--348, 2013.

\bibitem[Kim et~al.(2005)Kim, Singh, Lee, Lerch, Ad-Dab'bagh, MacDonald, Lee,
  Kim, and Evans]{Kim:NI2005:CLASP}
J.~S. Kim, V.~Singh, J.~K. Lee, J.~Lerch, Y.~Ad-Dab'bagh, D.~MacDonald, J.~M.
  Lee, S.~I. Kim, and A.~C. Evans, ``Automated 3-d extraction and evaluation of
  the inner and outer cortical surfaces using a laplacian map and partial
  volume effect classification,'' \emph{NeuroImage}, vol.~27, no.~1, pp.
  210--221, 2005.

\bibitem[Shattuck and Leahy(2002)]{Shattuck:MIA2002:BrainSuite}
D.~W. Shattuck and R.~M. Leahy, ``Brainsuite: an automated cortical surface
  identification tool,'' \emph{Medical Image Analysis}, vol.~6, no.~2, pp.
  129--142, 2002.

\bibitem[Kriegeskorte and Goebel(2001)]{Kriegeskorte:NI2001:BrainVoyage}
N.~Kriegeskorte and R.~Goebel, ``An efficient algorithm for topologically
  correct segmentation of the cortical sheet in anatomical mr volumes,''
  \emph{NeuroImage}, vol.~14, no.~2, pp. 329--346, 2001.

\bibitem[Fischl et~al.(1999)Fischl, Sereno, and Dale]{Fischl:NI1999:freesurfer}
B.~Fischl, M.~I. Sereno, and A.~M. Dale, ``Cortical surface-based analysis: Ii:
  inflation, flattening, and a surface-based coordinate system,''
  \emph{NeuroImage}, vol.~9, no.~2, pp. 195--207, 1999.

\bibitem[Henschel et~al.(2020)Henschel, Conjeti, Estrada, Diers, Fischl, and
  Reuter]{henschel2020:fastsurfer}
L.~Henschel, S.~Conjeti, S.~Estrada, K.~Diers, B.~Fischl, and M.~Reuter,
  ``Fastsurfer-a fast and accurate deep learning based neuroimaging pipeline,''
  \emph{NeuroImage}, p. 117012, 2020.

\bibitem[Rebsamen et~al.(2020)Rebsamen, Suter, Wiest, Reyes, and
  Rummel]{rebsamen2020brain}
M.~Rebsamen, Y.~Suter, R.~Wiest, M.~Reyes, and C.~Rummel, ``Brain morphometry
  estimation: From hours to seconds using deep learning,'' \emph{Frontiers in
  neurology}, vol.~11, p. 244, 2020.

\bibitem[He et~al.(2016)He, Zhang, Ren, and Sun]{he2016deep}
K.~He, X.~Zhang, S.~Ren, and J.~Sun, ``Deep residual learning for image
  recognition,'' in \emph{Proceedings of the IEEE conference on computer vision
  and pattern recognition}, 2016, pp. 770--778.

\bibitem[Jack~Jr et~al.(2008)Jack~Jr, Bernstein, Fox, Thompson, Alexander,
  Harvey, Borowski, Britson, L.~Whitwell, Ward, et~al.]{Jack2008:ADNI}
C.~R. Jack~Jr, M.~A. Bernstein, N.~C. Fox, P.~Thompson, G.~Alexander,
  D.~Harvey, B.~Borowski, P.~J. Britson, J.~L.~Whitwell, C.~Ward \emph{et~al.},
  ``The alzheimer's disease neuroimaging initiative (adni): Mri methods,''
  \emph{Journal of Magnetic Resonance Imaging}, vol.~27, no.~4, pp. 685--691,
  2008.

\bibitem[Rowe et~al.(2010)Rowe, Ellis, Rimajova, Bourgeat, Pike, Jones, Fripp,
  Tochon-Danguy, Morandeau, O'Keefe, et~al.]{rowe2010amyloid}
C.~C. Rowe, K.~A. Ellis, M.~Rimajova, P.~Bourgeat, K.~E. Pike, G.~Jones,
  J.~Fripp, H.~Tochon-Danguy, L.~Morandeau, G.~O'Keefe \emph{et~al.}, ``Amyloid
  imaging results from the australian imaging, biomarkers and lifestyle (aibl)
  study of aging,'' \emph{Neurobiology of aging}, vol.~31, no.~8, pp.
  1275--1283, 2010.

\bibitem[Desikan et~al.(2006)Desikan, S{\'e}gonne, Fischl, Quinn, Dickerson,
  Blacker, Buckner, Dale, Maguire, Hyman, et~al.]{desikan2006automated}
R.~S. Desikan, F.~S{\'e}gonne, B.~Fischl, B.~T. Quinn, B.~C. Dickerson,
  D.~Blacker, R.~L. Buckner, A.~M. Dale, R.~P. Maguire, B.~T. Hyman
  \emph{et~al.}, ``An automated labeling system for subdividing the human
  cerebral cortex on mri scans into gyral based regions of interest,''
  \emph{Neuroimage}, vol.~31, no.~3, pp. 968--980, 2006.

\bibitem[Hornik(1991)]{hornik1991approximation}
K.~Hornik, ``Approximation capabilities of multilayer feedforward networks,''
  \emph{Neural networks}, vol.~4, no.~2, pp. 251--257, 1991.

\bibitem[Huang et~al.(2019)Huang, Cheng, Bapna, Firat, Chen, Chen, Lee, Ngiam,
  Le, Wu, et~al.]{huang2019gpipe}
Y.~Huang, Y.~Cheng, A.~Bapna, O.~Firat, D.~Chen, M.~Chen, H.~Lee, J.~Ngiam,
  Q.~V. Le, Y.~Wu \emph{et~al.}, ``Gpipe: Efficient training of giant neural
  networks using pipeline parallelism,'' in \emph{Advances in neural
  information processing systems}, 2019, pp. 103--112.

\bibitem[Glorot and Bengio(2010)]{glorot2010understanding}
X.~Glorot and Y.~Bengio, ``Understanding the difficulty of training deep
  feedforward neural networks,'' in \emph{Proceedings of the thirteenth
  international conference on artificial intelligence and statistics}, 2010,
  pp. 249--256.

\bibitem[He and Sun(2015)]{he2015convolutional}
K.~He and J.~Sun, ``Convolutional neural networks at constrained time cost,''
  in \emph{Proceedings of the IEEE conference on computer vision and pattern
  recognition}, 2015, pp. 5353--5360.

\bibitem[Li et~al.(2018)Li, Xu, Taylor, Studer, and
  Goldstein]{li2018visualizing}
H.~Li, Z.~Xu, G.~Taylor, C.~Studer, and T.~Goldstein, ``Visualizing the loss
  landscape of neural nets,'' in \emph{Advances in Neural Information
  Processing Systems}, 2018, pp. 6389--6399.

\bibitem[Izmailov et~al.(2018)Izmailov, Podoprikhin, Garipov, Vetrov, and
  Wilson]{izmailov2018averaging}
P.~Izmailov, D.~Podoprikhin, T.~Garipov, D.~Vetrov, and A.~G. Wilson,
  ``Averaging weights leads to wider optima and better generalization,''
  \emph{arXiv preprint arXiv:1803.05407}, 2018.

\bibitem[Koo and Li(2016)]{koo2016guideline}
T.~K. Koo and M.~Y. Li, ``A guideline of selecting and reporting intraclass
  correlation coefficients for reliability research,'' \emph{Journal of
  chiropractic medicine}, vol.~15, no.~2, pp. 155--163, 2016.

\bibitem[Cicchetti(1994)]{cicchetti1994guidelines}
D.~V. Cicchetti, ``Guidelines, criteria, and rules of thumb for evaluating
  normed and standardized assessment instruments in psychology.''
  \emph{Psychological assessment}, vol.~6, no.~4, p. 284, 1994.

\end{thebibliography}

\end{document}